\documentclass{article}

\usepackage{PRIMEarxiv}
\usepackage{amssymb}
\usepackage[T1]{fontenc}
\usepackage{graphicx}
\usepackage{color}
\graphicspath{{images/}} 
\usepackage{color, soul}
\usepackage{microtype}
\soulregister\cite7
\soulregister\ref7
\soulregister\pageref7
\usepackage{multirow}
\usepackage{comment}
\usepackage{amssymb}
\usepackage{pifont}
\usepackage{amsmath} 
\usepackage{caption}
\usepackage{algorithmic}
\usepackage{algorithm}
\usepackage{booktabs}
\usepackage{subcaption}
\usepackage{dsfont}
\usepackage{url}

\title{GShield: Mitigating Poisoning Attacks in Federated Learning}

\author{
Sameera K. M.\\
Department of Computer Science and Engineering,\\ 
JAIN (Deemed-to-be University), Kochi,  India \\
\texttt{sameera.km@jainuniversity.ac.in} \\
\And
Serena Nicolazzo \\
Department of Science and Technological Innovation, \\
University of Eastern Piedmont, \\
V.le Teresa Michel, 11, Alessandria, Italy\\
\texttt{serena.nicolazzo@uniupo.it} \\
\And
Antonino Nocera\\
Department of Electrical, Computer and Biomedical Engineering,\\
University of Pavia, Italy\\
\texttt{antonino.nocera@unipv.it} \\
\And
Vinod P. \\
Department of Computer Applications, \\
Cochin University of Science \\
and Technology, India \\
\texttt{vinod.p@cusat.ac.in} \\
\And
Rafidha Rehiman K. A.\\
Department of Computer Applications, \\
Cochin University of Science \\
and Technology, India \\
\texttt{rafidharehimanka@cusat.ac.in} \\
}

\begin{document}

\maketitle

\date{December 2025}

\begin{abstract}
Federated Learning (FL) has recently emerged as a revolutionary approach to collaborative training Machine Learning models. In particular, it enables decentralized model training while preserving data privacy, but its distributed nature makes it highly vulnerable to a severe attack known as Data Poisoning. In such scenarios, malicious clients inject manipulated data into the training process, thereby degrading global model performance or causing targeted misclassification. In this paper, we present a novel defense mechanism called GShield, designed to detect and mitigate malicious and low-quality updates, especially under non-independent and identically distributed (non-IID) data scenarios. GShield operates by learning the distribution of benign gradients through clustering and Gaussian modeling during an initial round, enabling it to establish a reliable baseline of trusted client behavior. With this benign profile, GShield selectively aggregates only those updates that align with the expected gradient patterns, effectively isolating adversarial clients and preserving the integrity of the global model. An extensive experimental campaign demonstrates that our proposed defense significantly improves model robustness compared to the state-of-the-art methods while maintaining a high accuracy of performance across both tabular and image datasets. Furthermore, GShield improves the accuracy of the targeted class by $43\%$ to $65\%$ after detecting malicious and low-quality clients.
\end{abstract}


\section{Introduction} 

Federated Learning (FL, hereafter) is a new paradigm that enables the collaborative training of Machine Learning models across multiple decentralized devices or organizations while ensuring that raw data remains local and private \cite{mcmahan2017communication}. By transmitting only model updates rather than sensitive data, Federated Learning aims to preserve user privacy, reduce communication costs, and support compliance with data protection regulations. At the same time, it leverages the diversity and richness of distributed data sources to build models that are more robust, accurate, and applicable across a wide range of real-world applications.
While this design significantly enhances data privacy, it also raises new security vulnerabilities that are less severe in centralized learning. One of the most critical threats in FL is the poisoning attack
\cite{tolpegin2020data,sameera2024privacy}. In this kind of attack, malicious or compromised clients deliberately manipulate their local training data or model updates to corrupt the global learning process. Due to the distributed and often open nature of FL systems, the server typically lacks direct visibility into client data or training procedures, making it difficult to verify the integrity of individual updates. As a result, poisoned updates can be aggregated into the global model, leading to degraded performance or malicious behavior.

The challenge is particularly acute in realistic non-IID settings, where the natural heterogeneity of client data distributions causes benign updates to exhibit high variance. This variability significantly reduces the effectiveness of existing defense mechanisms, which often rely on simple distance-based outlier detection or robust aggregation rules that assume relatively homogeneous client behavior. As a result, poisoned updates can closely mimic legitimate ones, evade detection, and be aggregated into the global model, leading to degraded performance or targeted misclassification. These limitations motivate the need for a defense strategy that can accurately characterize benign client behavior under non-IID conditions and selectively filter malicious or low-quality contributions. 

To address this challenge, we propose GShield, a defense mechanism that learns the distribution of benign client gradients through clustering and Gaussian modeling during an initial trusted phase. By establishing a reliable baseline of expected gradient patterns, GShield enables the server to identify and isolate anomalous updates while preserving the contributions of honest clients, thereby maintaining both robustness and high model accuracy in adversarial federated environments.

GShield operates in two distinct phases. During an initial {\em Safe Round} phase, the system assumes the absence of adversarial clients and collects gradient updates from participating clients. By applying cosine-similarity-based clustering, GShield identifies the dominant, high-density cluster of updates, which is presumed to represent benign behavior. Statistical characteristics of this cluster, specifically the mean and variance of similarity scores, are aggregated over multiple rounds to construct a Gaussian model that serves as a baseline representation of trusted client updates. This benign profile captures natural variability while filtering out outliers and low-quality contributions.

In the subsequent {\em Anomaly Detection} phase, GShield evaluates incoming client updates by comparing their gradient similarity scores against the learned benign distribution. Updates that conform to this distribution are selectively aggregated, while those exhibiting significant deviation are excluded from the global model update. This selective aggregation mechanism enables GShield to effectively isolate malicious or unreliable clients, preventing poisoned updates from influencing the global model. As a result, GShield preserves both model accuracy and robustness, even in the presence of targeted label-flipping attacks and severe data heterogeneity.

In our experimental evaluation, we compare the proposed GShield approach against several well-established and widely adopted defenses, including FLAME \cite{nguyen2022flame}, Krum \cite{blanchard2017machine}, Median \cite{yin2018byzantine}, and Trimmed Mean (TMean) \cite{wang2025federated}. These methods are commonly used as baselines in the Federated Learning literature due to their demonstrated effectiveness across a wide range of attack models and threat scenarios~\cite{thein2024personalized, chelli2023fedguard, yang2025enhanced}. In line with this established evaluation practice, we include these representative techniques to ensure a fair, comprehensive, and consistent assessment of GShield's robustness and performance.

By grounding anomaly detection in a statistically learned benign baseline rather than rigid heuristics or global assumptions, GShield provides a principled and adaptive defense against targeted poisoning attacks in federated learning, ensuring reliable model convergence and improved resilience across diverse datasets and attack scenarios.

In summary, the main contributions of this paper are as follows:

\begin{itemize}
    \item We propose GShield, a novel server-side defense for Federated Learning that mitigates targeted data poisoning attacks without requiring prior knowledge of adversaries, clean auxiliary data, or assumptions on client data distributions.
    
    \item We introduce a statistically grounded approach that models benign client behavior through cosine-similarity-based clustering and Gaussian distribution learning, enabling reliable detection of malicious and low-quality updates under non-IID data.
    
    \item We validate GShield through comprehensive experiments on tabular and image datasets, demonstrating superior robustness and significant improvements in targeted attack mitigation compared to state-of-the-art defenses.
\end{itemize}

The outline of this paper is as follows. In Section \ref{sec:related}, we discuss the related papers in the state of the art. In Section \ref{sec:background}, we describe the concepts useful to understand our approach. Section \ref{sec:approach} presents our approach defense altogether the threat model. In Section \ref{sec:experiments}, we present the experiments carried out to test our defense. Finally, Section \ref{sec:conclusion} examines intriguing leads as future work and draws our conclusions.

\section{Related Works}
\label{sec:related}

Recently, robust aggregation methods have been employed during the server-side aggregation phase to enhance resilience against Poisoning Attacks. Traditional approaches like FedAvg \cite{mcmahan2017communication}, which aggregate client updates using a simple arithmetic mean, have been shown to be vulnerable to such attacks. To address this, various robust aggregation techniques have been proposed to identify and mitigate the influence of malicious updates. For instance, the Median method replaces the mean with the element-wise median of client updates \cite{yin2018byzantine}, while the Trimmed Mean method removes a fixed proportion of the smallest and largest values before averaging \cite{wang2025federated}. Norm Clipping constrains the impact of individual clients by bounding the norm of their updates to a predefined threshold \cite{sun2019really}. The Krum algorithm selects a single client update whose distance to the majority of other updates is minimal, aiming to reflect consensus \cite{blanchard2017machine}. Multi-Krum extends this by selecting and averaging the top-k closest updates, further enhancing robustness against outliers and adversarial contributions \cite{blanchard2017machine}.
In their research, the authors of \cite{guerraoui2018hidden} introduced Bulyan, a robust aggregation method that combines MultiKrum and trimmed-mean techniques to defend against poisoning attacks. However, a key limitation of Bulyan is its assumption of IID (independent and identically distributed) data, which does not hold in many federated learning scenarios. To address malicious behavior, Fung et al. proposed FoolsGold \cite{fung2020mitigating}, which filters out adversarial updates by analyzing the cosine similarity of gradients, assuming that malicious clients tend to produce similar updates. In \cite{pillutla2022robust}, the authors presented Robust Federated Aggregation (RFA), which aggregates model updates using a smoothed Weiszfeld algorithm to compute a weighted geometric median, thereby preserving client data privacy while enhancing robustness. Other notable approaches include FLAME \cite{nguyen2022flame}, ZeKoC \cite{chen2020zero}, FLTrust \cite{cao2022fltrust}, ShieldFL \cite{ma2022shieldfl}, Adaptive Model Averaging \cite{munoz2019byzantine}, Residual-based Reweighting \cite{fu2021attack}, SEAR \cite{zhao2021sear}, and FedGuard \cite{chelli2023fedguard}, each offering distinct mechanisms to mitigate the impact of adversarial or unreliable updates.

In our experimental campaign, we compared our GShield approach with FLAME \cite{nguyen2022flame}, Krum \cite{blanchard2017machine}, Median \cite{yin2018byzantine}, and TMean \cite{wang2025federated} methods. Prior works typically compare their proposed approaches with these techniques due to their effectiveness and popularity across various attack scenarios~\cite{thein2024personalized,chelli2023fedguard,yang2025enhanced}. Following this convention, we also include these representative methods in our experimental comparison to provide a fair and consistent evaluation of our proposed GShield.

\section{Background}
\label{sec:background}
In this section, we introduce several foundational concepts that support our approach. Specifically, we explore the main characteristics and categories of Federated Learning (FL) and examine the main features of poisoning attacks in this context.

Table \ref{tab:SystemSymbols} summarizes the acronyms used in this paper.

\begin{table}
\centering
  \caption{Summary of the acronyms used in the paper}
  \begin{tabular}{ll}
\hline
    \textbf{Symbol} & \textbf{Description}\\
\hline
    BA & Backdoor Attack\\
    DNN & Deep Neural Network\\
    DL & Deep Learning\\
    FL & Federated Learning\\
    GM & Global Model\\
    LFA & Label Flipping Attack \\
    LM & Local Model\\
    MIA & Membership Inference Attack\\
    ML & Machine Learning\\
    PM & Poisoned Model\\
\hline
\end{tabular}
\label{tab:SystemSymbols}
\end{table}

\subsection{Federated Learning}

Federated Learning (FL, hereafter) is a distributed approach developed to address optimization problems in an environment where a central server coordinates the learning process across a large number of decentralized devices. The devices, known as {\em clients} or {\em workers}, collaboratively train a Global Model (GM) by processing their local data samples and actively preserving data privacy by avoiding the exchange of raw data. We assume that each client \( P_i \), for all \( i \in [1, N] \), possesses a local training dataset \( \mathcal{D}_i \). These datasets are mutually disjoint and follow a non-identically independent distributed (non-IID) pattern, i.e., \( \mathcal{D}_i \cap \mathcal{D}_j = \emptyset \) for all \( i, j \in [1, N] \) such that \( P_i \neq P_j \). The FL objective seeks to minimize a global loss function \( f(w) \), defined as the average of local objectives across \( N \) clients:

\begin{equation}
\min_{w} f(w) = \frac{1}{N} \sum_{i=1}^{N} F_i(w)
\label{eq:functionloss}
\end{equation}

Here, \( w \) denotes the parameters of the shared global model. For simplicity, we represent both the input data and the model parameters as vectors. The local objective \( F_i(w) \) for client \( i \) typically corresponds to the empirical risk minimization, defined over its local data distribution \( \mathcal{D}_i \) as:

\begin{equation}
F_i(w) = \mathbb{E}_{x_i \sim \mathcal{D}_i} \left[ \ell_i(w; x_i) \right]
\label{eq:functionFi}
\end{equation}

\noindent where \( \ell_i() \) is the local loss function and \( x_i \) denotes the training samples.
In this work, we consider \( F_i(w) \) to be a non-convex function, which is optimized locally by each client using stochastic solvers such as Stochastic Gradient Descent (SGD). During each communication round, clients receive the current global model \( w^t \) from the Aggregator Server, and perform \( E \) epochs of local updates as follows: 

\begin{equation}
w_{i,E}^{t} = w_{i,0}^{t} - \eta \sum_{e=0}^{E-1} \nabla F_i\left(w_{i,e}^{t}; \mathcal{B}_{i,e} \right)
\label{eq:functionwi}
\end{equation}

\noindent where \( \eta \) is the local learning rate, and \( \mathcal{B}_{i,e} \) denotes the mini-batch sampled at epoch \( e \) on client \( i \). The updated model \( w_i^{t,E} \) is then sent to the central server for aggregation.

\begin{figure*}
    \centering
    \includegraphics[width=0.6\linewidth]{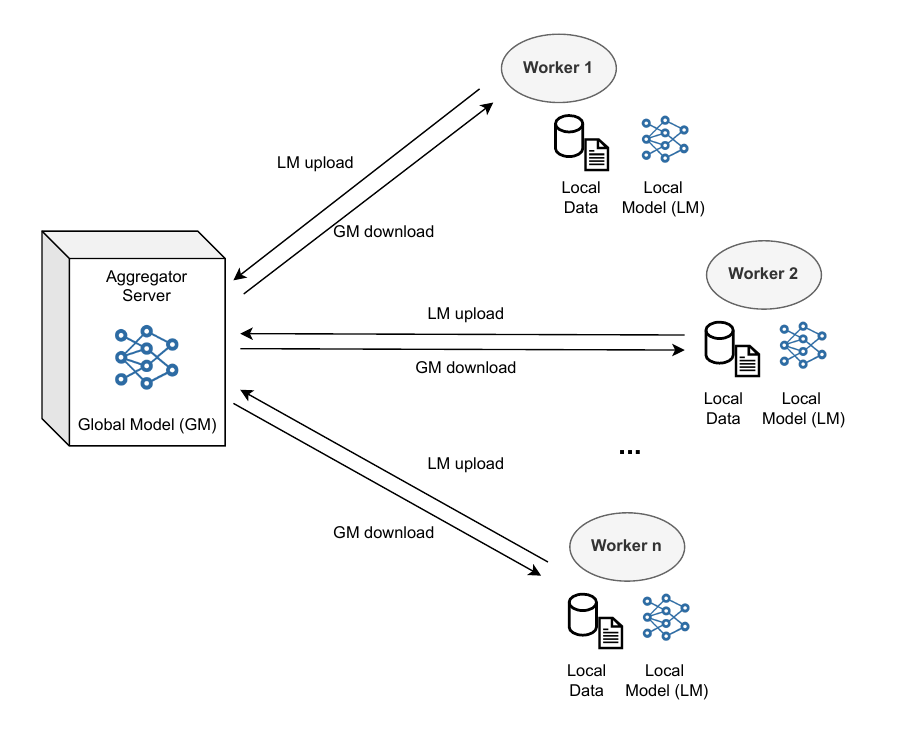}
    \caption{A general Federated Learning architecture}
    \label{fig:FL}
\end{figure*}

Figure \ref{fig:FL} shows a general architecture of FL.

\subsection{Poisoning Attacks on Federated Learning}

Poisoning attacks aim to subvert the learning process by introducing malicious perturbations, either through crafted data samples or manipulated local model updates. Their objective is to degrade the performance or integrity of the global model. In Federated Learning (FL), clients collaboratively train a shared model by periodically transmitting local updates to a central server, which aggregates them to refine the global model. Due to the decentralized and trust-assumed nature of FL, a compromised client can inject adversarial updates into the aggregation process. These poisoned contributions propagate through the model, leading to convergence instability, biased predictions, or a significant drop in overall model accuracy.
There exist two main categories of Poisoning Attacks, each exploiting different aspects of FL's decentralized architecture, and requiring specific defense strategies, namely:

\begin{itemize}
    \item \textbf{Data Poisoning Attacks}, in which the training data used by local clients is targeted.
    \item \textbf{Model Poisoning Attacks}, in which the local model updates sent to the server are manipulated.
\end{itemize}

\begin{figure*}[ht]
    \centering
    \includegraphics[width=0.6\linewidth]{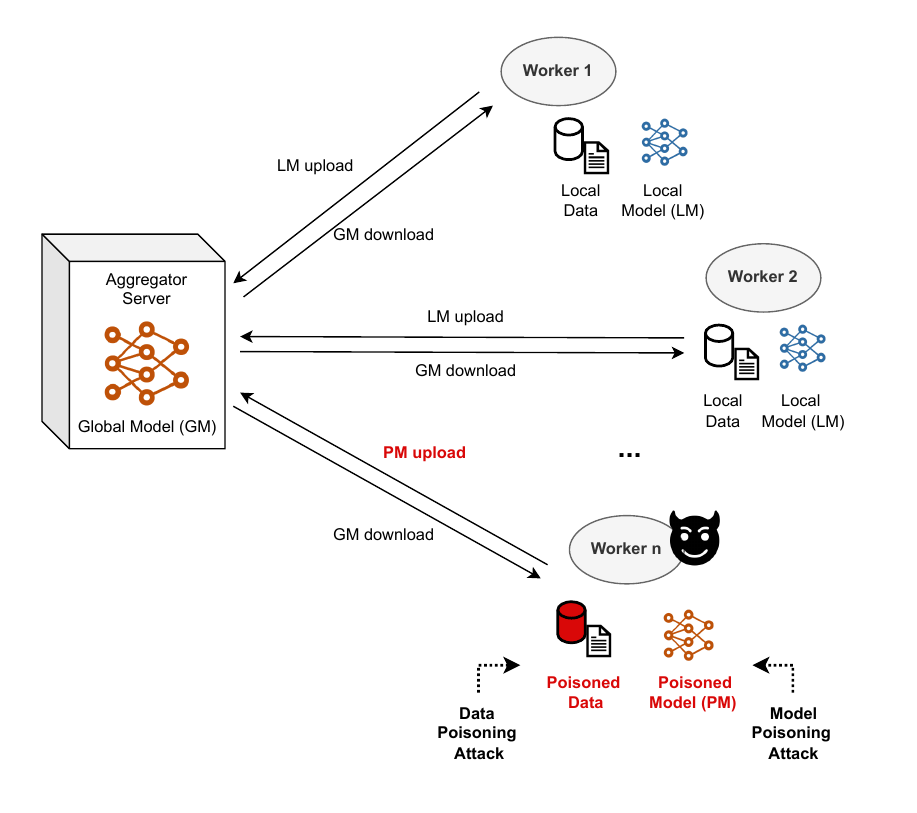}
    \caption{Example of Poisoning Attacks in FL}
    \label{fig:poisonAttackExample}
\end{figure*}

Figure \ref{fig:poisonAttackExample} illustrates the scheme of the two categories of Poisoning Attack in FL. During the FL training process, an attacker can execute a Poisoning Attack by modifying the client's local data or directly modifying the local model
parameters.

Beyond the specific techniques used, both Model and Data Poisoning Attacks can be further categorized based on the attacker's intent. Broadly, these intents fall into two categories depending on the scope and specificity of the adversary's objective: {\em(i) Untargeted Poisoning}, which aims at degrading the overall performance of the global model across all tasks, or {\em(ii) Targeted Poisoning}, which selectively impairing the model's behavior on particular inputs or tasks. Additionally, there exists also a third category known as {\em Semi-targeted Poisoning} in which the adversary designates a specific source class (denoted as $C_S$) and aims to corrupt the GM such that instances from this class are misclassified as a different, attacker-chosen target class ($C^*$). The general formulation is given by:

$$\begin{aligned} \arg \max _m f(x)_m = {\left\{ \begin{array}{ll} C^* &  \text {if } y = C_S \\ y &  \text {otherwise} \end{array}\right. } \end{aligned}$$

\subsubsection{Data Poisoning Attacks}
These attacks target the training data used by local clients, contaminating the local dataset during training. The can be categorized in the following types: {\em (i)} Label Flipping, {\em (ii)} Data Injection, and {\em (iii)} Backdoor techniques. 

In {\em Label Flipping Attack}~(LFA), the attacker flips labels of training samples (e.g., labeling a ``fish'' as a ``dog''). Label-flipping attacks can be classified as clean-label and dirty-label variants. In clean-label scenarios, the adversary is constrained from modifying class labels, typically due to a certification mechanism that ensures label integrity and limits perceptible input manipulation. This attack can be perpetrated by modifying
the samples in the training set and, for example, adding noise to the training data. Conversely, dirty-label attacks allow the adversary to inject training samples with intentionally incorrect labels—distinct from their true class—thereby inducing misclassifications toward a specific target class. 
In {\em Data injection Attacks}, a malicious actor tries to modify specific data samples within a dataset, injecting anomalous or irrelevant data points to skew local model updates.
Finally, in {\em Backdoor Attack} involves an adversary embedding a covert mechanism into the model during training, enabling controlled manipulation of its predictions. This is achieved by associating specific, often imperceptible input patterns—known as triggers—with a target misclassification. The attack introduces a malicious perturbation $\Delta$ that activates only when the trigger is present, causing the model to misclassify the input while preserving standard behavior on clean, trigger-free data \cite{liu2022backdoor,nicolazzo2025secure}.

\begin{equation}
    F(X) = y_i, \quad \forall X \sim \mathcal{D}_{\text{clean}}
\end{equation}

\begin{equation}
    F(X + \Delta) = y_t, \quad \forall (X+ \Delta) \sim \mathcal{D}_{\text{backdoor}}
\end{equation}

\noindent
where $X$ represents the clean input data; $\Delta$ is the trigger pattern introduced by the attacker; $y_i$ denotes the true label of the clean input; and $y_t$ represents the attacker's target label. The clean data is drawn from the distribution $\mathcal{D}_{\text{clean}}$, whereas the manipulated (backdoored) data comes from the distribution $\mathcal{D}_{\text{backdoor}}$. 

\subsubsection{Model Poisoning Attacks}
These kinds of attacks manipulate the local model updates sent to the aggregator server.
We can organize this type of attack into three categories \cite{sameera2025federated}, depending on how the adversary creates the models, namely: {\em (i)} Model Replacement, {\em (ii)} Optimization-based method, and {\em (iii)} Training Rule Manipulation.

In {\em Model Replacement Attack}, the attacker substitutes the models or gradients with crafted ones, achieved through random weights generation or gradient manipulation. In the second category of {\em Optimization-based method}, the adversary subtly alters the local training process (for instance, using different learning rates or objectives) to distort updates without obvious signs. Finally, in {\em Training Rule Manipulation}, the training rule is altered to introduce poison into the global model.

\section{Overview of GShield}
\label{sec:approach}

This section presents an overview of the proposed robust defense framework called GShield, designed to mitigate data poisoning attacks in the FL framework. We begin by examining the threat model related to targeted LFA in FL. Next, we introduce the GShield framework, highlighting its main components and the algorithms it employs. This is followed by a theoretical analysis of its effectiveness in mitigating poisoning attacks.

\subsection{Threat Model}
\begin{figure*}
    \centering
    \includegraphics[width=0.8\linewidth]{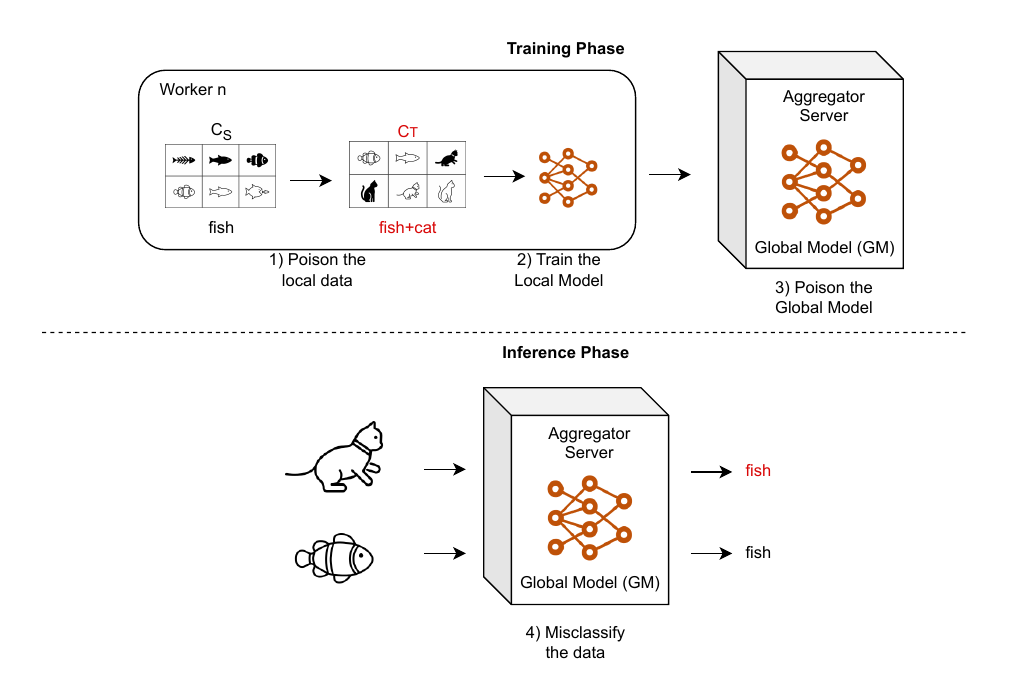}
    \caption{Our threat scenario}
    \label{fig:attack}
\end{figure*}

Targeted Label Flipping Attacks in FL are a type of Poisoning Attack where malicious clients attempt to corrupt the Global Model (GM) in such a way that specific inputs are misclassified into a target class chosen by the attacker. In particular, the adversary's goal is to manipulate the local training data updates to force the GM to misclassify specific inputs (from the source class) as belonging to a specific target label, without significantly degrading overall model performance. This is different from untargeted attacks, which aim to reduce the overall accuracy or cause arbitrary misclassifications. We assume the Aggregator Server to be honest and non-compromised,
and the attacker (that can be more than one entity) to have no control over the the Aggregator or the honest peers.
In particular, as visible in Figure \ref{fig:attack}, the attackers poison their local training
data by flipping the labels of training examples of a source class $C_S$ to a
target class $C_T \in C$ while keeping the input data features unchanged. In brief, the steps of the attack are the following:
\begin{itemize}
    \item Every instance with the source label $C_S$ is changed to the target label $C_T$, to introduce intentional label noise into the training data.
    \item The client trains the local model using this poisoned dataset for a fixed number of epochs.
    \item The poisoned model update is sent to the Aggregator server during the standard federated aggregation step without being detected by the aggregation methods.
    \item Over multiple rounds, the GM starts to misclassify inputs from the source class as the target class.
\end{itemize}

\subsection{Defense Approach Description}

This section presents an overview of the proposed robust defense framework called GShield, designed to mitigate label-flipping poisoning attacks in the FL system framework. Our proposed GShield functions without prior knowledge of attacker proportions, auxiliary datasets, or assumptions about the data distribution. It is deployed on the server side, aiming primarily to identify malicious clients and low-quality models present in the heterogeneous environment.

\begin{algorithm}[ht!]
\caption{\texttt{GShield}: Defense against targeted poisoning attack }
\label{alg:GShield_overall}
\begin{algorithmic}[1]
\REQUIRE Global Model $\omega^0$, dataset $D_i$ for each client $i$, list of compromised clients $Adv$
\ENSURE Final Global Model $\omega^T$ after $T$ rounds
\FOR{$t = 1$ to $T$}
    \STATE $S^t \leftarrow$ Randomly select $K$ clients
    \STATE Server sends $w^{t-1}$ to all clients in $S^t$
    \FORALL{client $i \in S^t$ \textbf{in parallel}}
        \IF{$i \in Adv$}
            \STATE Apply label flipping attack on $D_i$
        \ENDIF
        \STATE $g_i^t \leftarrow$  Compute gradients using SGD on local data [$\nabla \mathcal{L}(w_i^{t-1}; D_i)$]
        \STATE Client $i$ sends $g_i^t$ to the server
    \ENDFOR
    \\ /*Detect benign clients using gradients*/
    \STATE $\mathcal{B}^t \leftarrow$ \texttt{BenignClientSelection}($\{g_i^t\}_{i \in S^t}$) 
 \STATE $w^t \leftarrow$ Aggregate model from benign clients in $\mathcal{B}^t$

\ENDFOR
\RETURN $\omega^T$
\end{algorithmic}
\end{algorithm}

GShield is a server-side defense mechanism designed to detect and mitigate the influence of malicious and low-quality model updates in federated learning. It consists of two main phases:
\begin{itemize}
    \item \textbf{Safe Phase}. In the initial training rounds belonging to this phase GShield aims to learn and model benign client behavior by building a statistical baseline of trustworthy gradient updates.
    \item \textbf{Anomaly Detection Phase}. During this phase, the system aims to detect and filter malicious or low-quality updates by comparing them against the learned benign baseline.
\end{itemize}

GShield begins with an initial phase, termed the \textit{Safe Round}, during which it operates under the assumption that no adversarial clients are present. In this phase, GShield collects gradient updates from participating clients and applies clustering techniques to partition them into distinct groups. The dominant group, identified as the \textit{positive cluster}, is presumed to comprise benign clients. In contrast, the remaining group, containing outliers and low-quality updates, is designated as the \textit{negative cluster}. GShield then computes statistical measures, including the mean and standard deviation of the updates in the \textit{positive cluster}, to construct a Gaussian Mixture Model (GMM). The global model is updated exclusively by aggregating the contributions from the \textit{positive cluster}. This process is repeated for each round of the \textit{Safe Round} phase, establishing a robust baseline that facilitates accurate anomaly detection and mitigation in subsequent training stages.

\begin{figure*}
    \centering
    \includegraphics[width=0.8\linewidth]{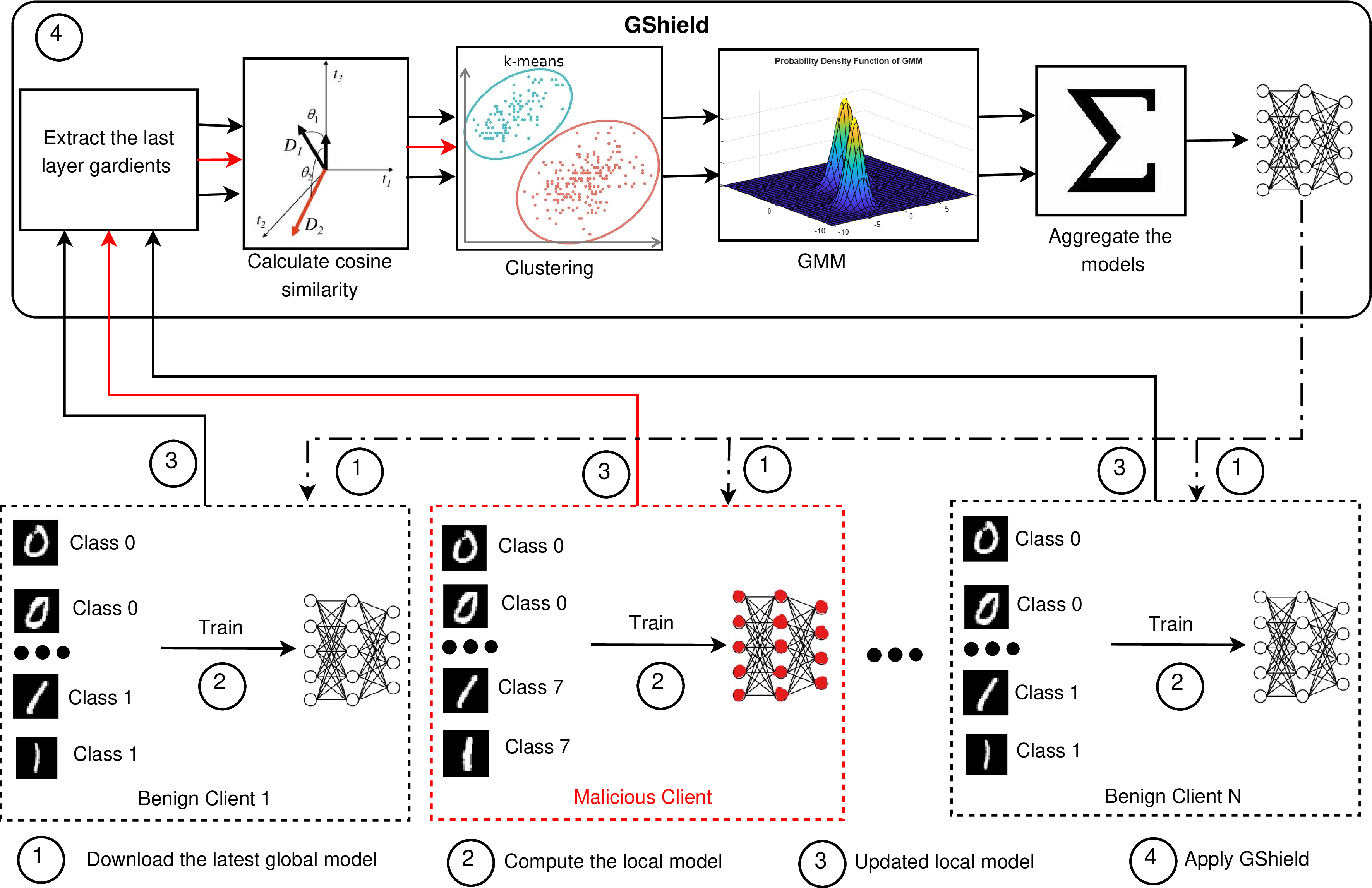}
    \caption{Proposed GShield receives local updates from clients and generates a new global model that eliminates poisoned local updates.}
    \label{fig:GShield_Architecture}
\end{figure*}

In the subsequent rounds, known as the \textit{Anomaly Detection Phase}, GShield uses the previously learned GMM to assess incoming client updates. It filters out suspicious or inconsistent gradients by comparing them to the benign distribution established during the \textit{Safe Round} phase. Only updates that conform to the expected benign behavior are aggregated. This selective aggregation process enables GShield to effectively identify and mitigate the influence of malicious or low-quality clients, thus preserving the integrity and performance of the global model throughout the training process. Figure~\ref{fig:GShield_Architecture} illustrates the overall workflow of GShield, particularly highlighting its operations during the \textit{Anomaly Detection Phase}.

\begin{algorithm}[h]
\caption{\texttt{BenignClientSelection}}
\label{alg:GShield}
\begin{algorithmic}[1]
\REQUIRE Local Model $\omega^0$, current round $t$, safe round $T_{\text{safe}}$, threshold $ z_{\alpha} $
\ENSURE Benign model $\mathcal{B}^t$

    \FORALL{client $i \in S^t$ }
        \STATE $g_i^{t,L} \leftarrow$ Extract last-layer gradient 
        \STATE Stack gradients $\{g_i^{t,L}\}$ into matrix $G^t$
    \ENDFOR
    \STATE $ Cos^t\leftarrow$ Compute the cosine similarity $G^t$ 
    \STATE  $(C_1, C_2) \leftarrow \text{KMeansClustering}(Cos^t, \text{num\_clusters}=2)$
        \STATE $\mathcal{C}_{\text{max}} \leftarrow$ Identify largest density cluster in $C_1,C_2$ 
        
    \IF{$t \leq T_{\text{safe}}$}
        \STATE $\mathcal{B}^t \leftarrow$ clients in $\mathcal{C}_{\text{max}}$
        \STATE $\mu_{safe}, \sigma_{safe} \leftarrow$ mean $\mu_t$ and standard deviation of $Cos^t$ of $\mathcal{B}^t$ 
        \IF{$t = T_{\text{safe}}$}
           \STATE $\mu, \sigma \leftarrow$ Average of $\mu_{safe}, \sigma_{safe}$
            \STATE Fit Gaussian model $\mathcal{N}(\mu, \sigma^2)$
        \ENDIF
    \ELSE
        \STATE $\mathcal{B}^t \leftarrow \emptyset$
        \FORALL{client $i \in S^t$}
            \STATE Compute cosine similarity score $\text{sim}_i$
            \STATE Compute Mahalanobis distance $d_i = \frac{|\text{sim}_i - \mu|}{\sigma}$
            \IF{$d_i \leq z_{\alpha}$}
                \STATE $\mathcal{B}^t \leftarrow \mathcal{B}^t \cup \{i\}$ \COMMENT{Mark client as benign}
            \ENDIF
        \ENDFOR
    \ENDIF


\RETURN $\mathcal{B}^t$
\end{algorithmic}
\end{algorithm}
We outline our method in Algorithm \ref{alg:GShield_overall}. At the beginning of round \( t \), the server randomly selects a set \( S^t \) of \( K \) clients and sends the current global model \( w^{t-1} \) to all clients in \( S^t \). Each selected client \( i \in S^t \) then performs local training on its dataset \( D_i \). If client \( i \) is an adversary (denoted as \( Adv \)), it applies a label-flipping attack to its dataset \( D_i \). After completing the local training, each client computes its gradient \( g_i^t = \nabla \mathcal{L}(w^{t-1}; D_i) \) using stochastic gradient descent and returns this gradient to the server. 

Upon receiving the \( K \) local gradients, the server implements a benign client detection procedure to identify which clients are likely behaving reliably. The server then aggregates the updates from these trustworthy clients to create the new global model \( w^t \), while isolating any suspected malicious or low-quality clients to maintain the integrity of the global model.

\noindent
The proposed \textit{Benign Client Selection} procedure, presented in Algorithm~\ref{alg:GShield}, operates in two distinct phases. In each round $t$, the server first collects the last-layer gradients $g_i^{t,L}$ from every participating client $i \in S^t$ and constructs a gradient matrix $G^t$. The server then computes the cosine similarity $Cos^t$ across these gradient vectors and applies $k$-means clustering with $k = 2$ to partition the clients into two clusters. The largest, densest cluster, denoted as $\mathcal{C}_\mathit{max}$, is presumed to comprise predominantly benign clients.




During the \textit{Safe Round} phase ($t \leq T_\mathit{safe}$), the server treats all clients in $\mathcal{C}_\mathit{max}$ as benign and records their average cosine similarity $\mu_t$ and standard deviation $\sigma_t$. Upon reaching $t = T_\mathit{safe}$, it aggregates these statistics across the safe rounds to obtain global estimates of the benign distribution, $\mu$ and $\sigma$, and fits a corresponding Gaussian model $\mathcal{N}(\mu, \sigma^2)$.

In the \textit{Anomaly Detection Phase} ($t > T_\mathit{safe}$), the server evaluates new client updates by computing their cosine similarity scores $\mathit{sim}_i$ and calculating the associated Mahalanobis distance.

\begin{equation}
    d_i = \frac{\lvert \mathit{sim}_i - \mu \rvert}{\sigma}.
\end{equation}
Clients with $d_i \leq z_\alpha$ are classified as benign and added to $\mathcal{B}^t$. This procedure allows the server to reliably filter out malicious or low-quality clients, thereby ensuring robust and secure federated learning throughout the training process.

\section{Experiments}
\label{sec:experiments}

In this section, we evaluate the performance of GShield compared to other existing approaches to assess its effectiveness and robustness. Specifically, we detail the experiment setup, including the employed datasets and the evaluation metrics, followed by a comprehensive presentation and critical analysis of the obtained results.

\subsection{Datasets and Implementation Details}

We extensively evaluated our proposed GShield method using publicly available benchmark datasets commonly adopted in this domain, encompassing both tabular and image-based modalities. For the tabular data, we considered two widely used network intrusion datasets: CIC-Darknet2020 and CSE-CIC-IDS2018~\cite{hernandez2025intrusion}. To assess the generalizability of our approach beyond network traffic, we also included two image-based benchmarks: FashionMNIST and EMNIST-Digits. We configured FL with a total number of clients of $50$, where $50\%$ of the participants are considered for the aggregation. To evaluate the robustness and performance of the proposed GShield, we examined varying  Poisoned Malicious Ratios (PMR), with $5\%$, $10\%$, $15\%$, $20\%$, and $25\%$ of the total clients configured as malicious participants. To control the degree of heterogeneity, we employed a Dirichlet distribution \cite{hsu2019measuring} with a concentration parameter of $\alpha = 0.5$ to generate non-IID training data for the clients.

\textbf{System and machine setup}. 
We conducted our experiments on an Ubuntu 22.04 LTS system equipped with an Intel Core i9 processor, 32 GB of RAM, and an NVIDIA Quadro P2000 GPU with 5 GB of GDDR5X memory. Our FL implementation was developed using the PyTorch framework, where clients were trained sequentially and independently in each global training round.

\subsection{Evaluation Metrics}
To assess the performance of our defense strategy, we mainly employed the following evaluation metrics:

\begin{itemize}

     \item \textbf{Source Class Recall}, which measures the proportion of correctly classified samples from the adversary's chosen source class among all samples that truly belong to that class.
     $$
    \text{SRecall} = \frac{1}{N_{\text{src}}} \sum_{i=1}^{N_{\text{src}}} (y_i = \hat{y}_i)
    $$
    where $N_{\text{src}}$ is the total number of instances belonging to the source class, $y_i$ is the true label of the $i$-th instance, and $\hat{y}_i$ is the predicted label.
     

     \item \textbf{F1-score}, that is, the harmonic mean of precision and recall, ranging between 0 and 1.
     $$\text{F1Score} = 2 \cdot \frac{\text{P} \cdot \text{R}}{\text{P} + \text{R}}$$
     Precision $(P)$ indicates the proportion of correctly predicted positive samples among all samples predicted as positive, while Recall $(R)$ represents the proportion of correctly predicted positive samples among all actual positive samples.

\end{itemize}

\subsection{Performance Analysis on Baseline Model}

\begin{figure*}[h!]
\centering
\begin{subfigure}{0.45\textwidth}
    \includegraphics[width=0.9\linewidth]{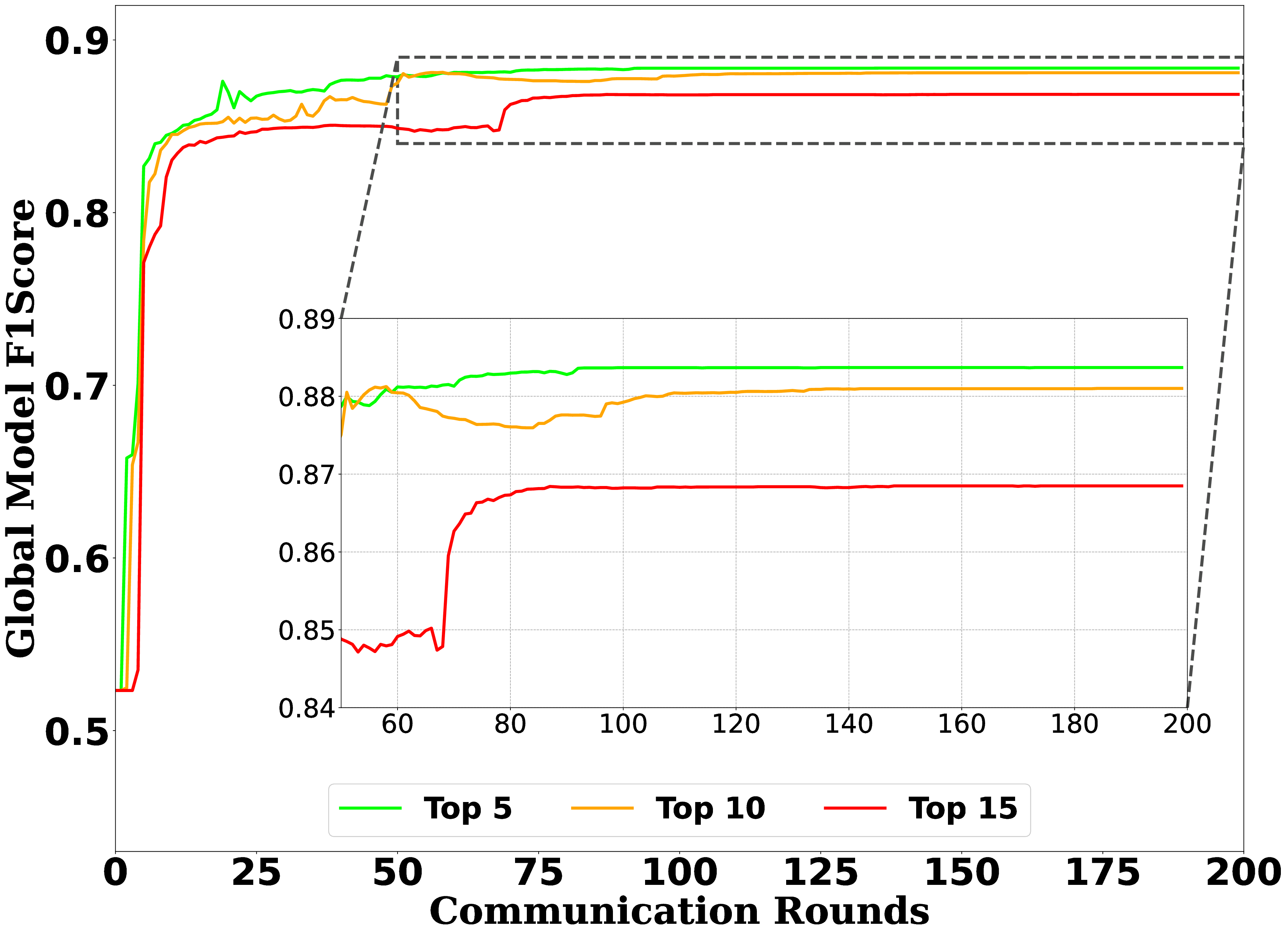} 
    \caption{CIC-Darknet2020}
    \label{fig:cicdarknet_noniid}
\end{subfigure}
\hfill
\begin{subfigure}{0.45\textwidth}
    \includegraphics[width=0.9\linewidth]{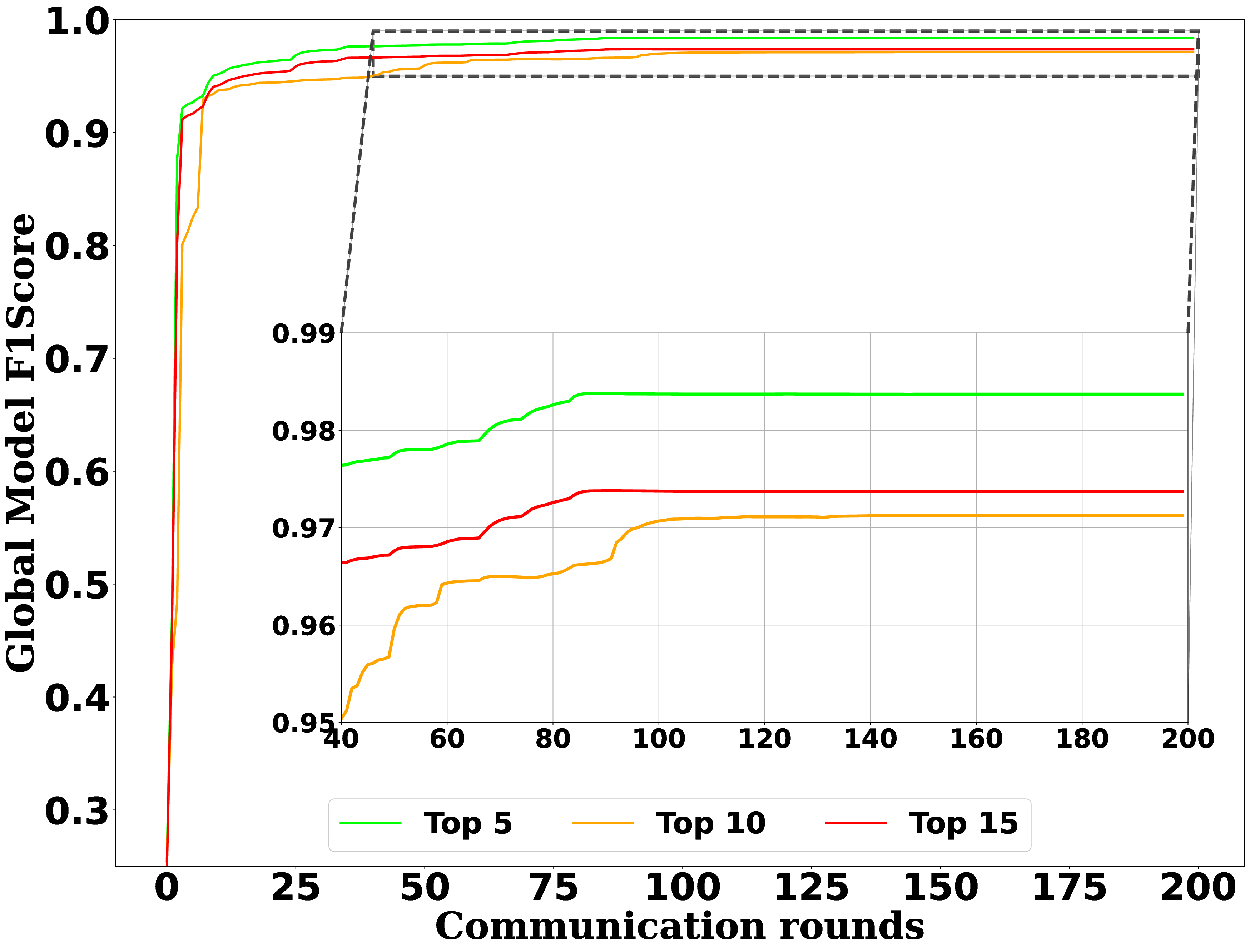} 
    \caption{CIC-IDS2018}
    \label{fig:cicids2018}
\end{subfigure}
\\
\begin{subfigure}{0.45\textwidth}
    \includegraphics[width=0.9\linewidth]{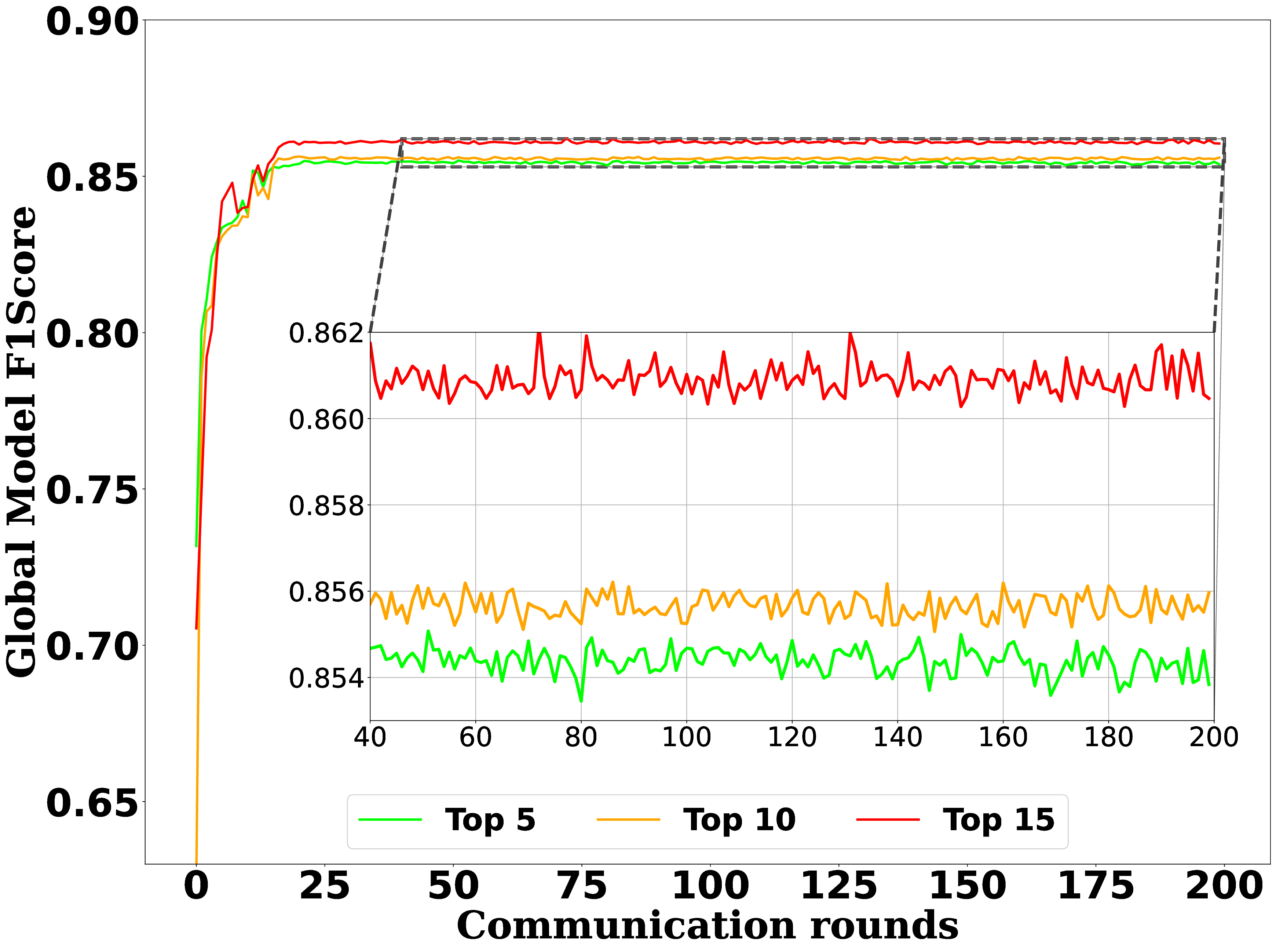} 
    \caption{FashionMNIST}
    \label{fig:FashionMNIST}
\end{subfigure}
\hfill
\begin{subfigure}{0.45\textwidth}
    \includegraphics[width=0.9\linewidth]{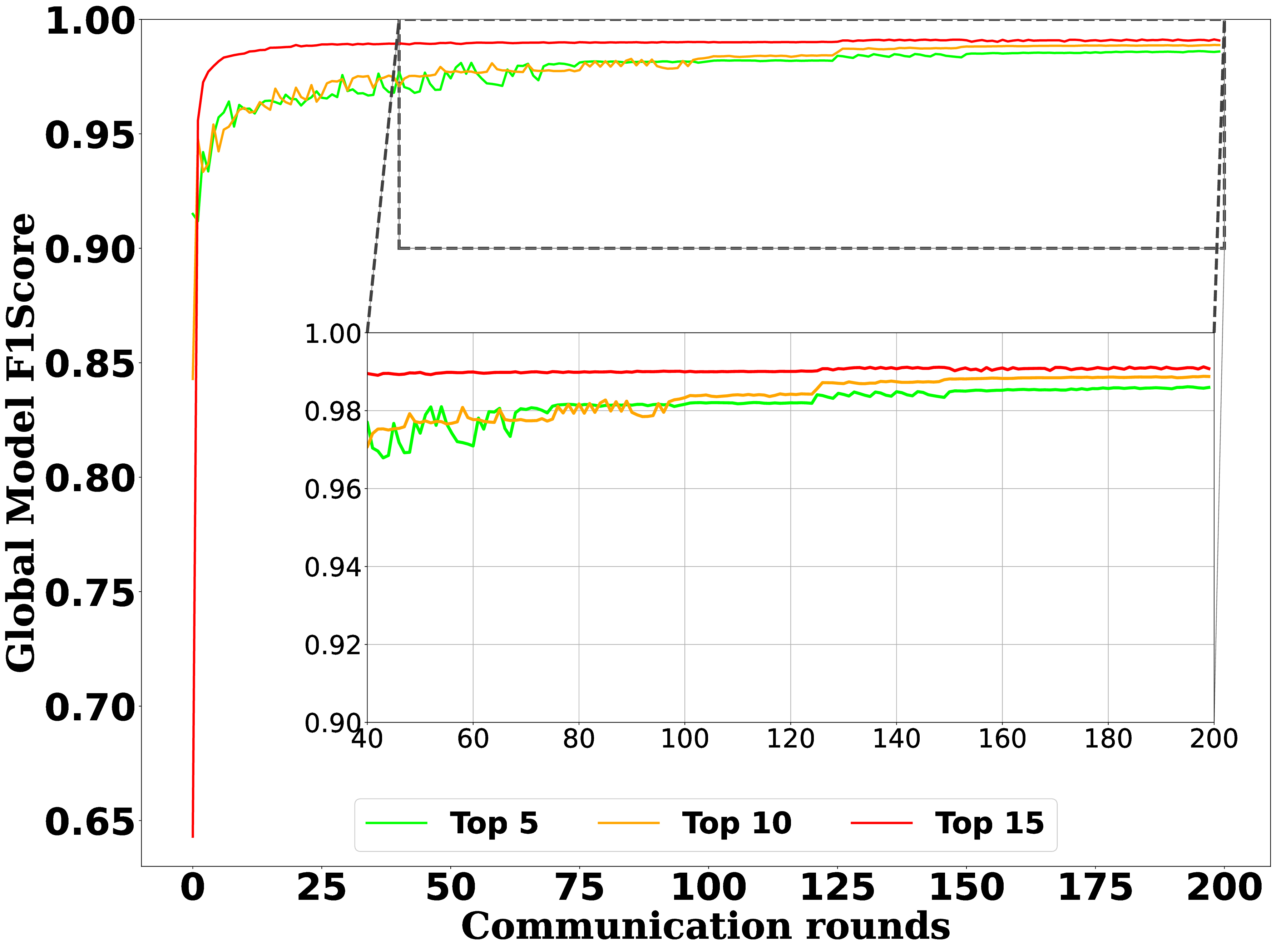} 
    \caption{EMNIST}
    \label{fig:FENMNIST}
\end{subfigure}
\hfill
\caption{Baseline.}
\label{fig:baseline_scenraio}
\end{figure*}

We conducted baseline evaluations under a non-poisoned scenario to analyze the F1 score of our FL configuration. Initially, the model was trained for $200$ communication rounds. From these results, we determined the optimal number of communication rounds for subsequent experiments. Figure \ref{fig:baseline_scenraio} illustrates the performance of the global model in F1-score matrices across these communication rounds, with aggregation performed using the top-k participants. 

We observed that the model started to converge after approximately $50$ communication rounds for both the CIC-Darknet2020 and CSE-CIC-IDS2018 datasets. Similarly, the model reached convergence in approximately $50$ rounds for both the FashionMNIST and EMNIST datasets. Consequently, we set the number of communication rounds to $100$ for tabular datasets and $50$ for image-based datasets in all subsequent experiments.

\begin{table*}[h!]
\centering
\caption{Comparison of proposed methods under different safe rounds on the tabular dataset. Bold indicates best results.}
\label{tab:tabulardataset}
\renewcommand{\arraystretch}{1.2}
\resizebox{\textwidth}{!}{
\begin{tabular}{cccccccccccc}
\hline
\multirow{2}{*}{$SafeRound\downarrow$}&\multirow{2}{*}{$PMR\downarrow$}  
& \multicolumn{2}{c}{Krum} 
& \multicolumn{2}{c}{TMean} 
& \multicolumn{2}{c}{Median} 
& \multicolumn{2}{c}{FLAME} 
& \multicolumn{2}{c}{GShield} \\
\cline{3-12}
&& F1Score & SRecall & F1Score & SRecall & F1Score & SRecall & F1Score & SRecall & F1Score & SRecall \\
\hline  
\multicolumn{12}{c}{\textbf{\textit{CIC-Darknet2020}}} \\
\hline
\multirow{5}{*}{5}&5\% &  0.7890 & 36.054 & 0.7144 &  14.582& 0.7313 &35.313 &0.8366 &36.213 & 0.8747& 79.466\\

&10\% &0.8399   & 60.223 & 0.8373 & 76.016 & 0.8403 &79.935 & 0.8448& 74.751&0.8780 & 88.364\\

&15\% &  0.8191 & 30.076 &0.8244  &27.157  & 0.7884 & 53.184&0.8031 &44.598 & 0.8662&88.94 \\

&20\% & 0.7253  &1.223  & 0.7162 &1.094  & 0.7191 &1.36&0.7328 &1.166& 0.8737& 83.247\\

&25\% &  0.7108 & 0.224 & 0.7012 & 0.201 &0.7076  &0 &0.7207 & 0.216&0.8761 &76.848 \\
\hline
\multirow{5}{*}{10}&5\% & .8373  & 63.559 & 0.7428 & 18.518 & 0.7152 &60.094 &0.7915 &37.891 & 0.8769&79.728 \\

&10\% &  0.8458 &  69.443&  0.8399& 75.737 & 0.8372 &79.636 &0.8392 &74.229 & 0.8811&87.7 \\

&15\% &  0.8078 &32.988  & 0.7997 & 31.724 & 0.8259 & 52.264& 0.8220& 45.472& 0.8687&87.933 \\

&20\% & 0.7364  & 2.96 & 0.7201 & 1.937 & 0.7171 & 2.379& 0.7291& 2.542&0.8688 &83.573 \\

&25\% &  0.7223 & 0.879 & 0.7075 & 0.394 & 0.7013 &0.24 &0.7085 & 0.72&0.8735 &75.24 \\
\hline
\multirow{5}{*}{15}&5\% & 0.8377 &63.753  & 0.7592 &  69.41& 0.7154 &24.117& 0.7956&39.895 &0.8781 & 81.409\\

&10\% & 0.8447  &  68.65&  0.8164&40.176  &  0.8369&75.21 &0.8391 & 74.092& 0.8799& 87.611\\

&15\% & 0.8094  &35.156  &  0.8113& 38.155 & 0.8271 & 53.99& 0.8237&48.013 & 0.8735& 87.262\\

&20\% & 0.7383  & 4.866 & 0.7237 &3.687  & 0.7241 & 5.445& 0.7327& 4.157& 0.8718&84.595 \\

&25\% &  0.7261 &2.381  &0.7115  & 1.595 &0.6997  & 0.492&0.7119 & 1.792&0.8642 & 62.256\\
\hline
\multicolumn{12}{c}{\textbf{\textit{CSE-CIC-IDS2018}}} \\
\hline
\multirow{5}{*}{5}&5\% 
&0.974 & 86.314 & 0.9729 & 84.237&  0.9714 &67.756 &0.9765 &85.338 &0.9821  & 91.094\\

&10\%& 0.960  & 70.163 &0.9676&  71.057 & 0.9530&62.979 &0.97147 &72.567 & 0.9839 &90.128 \\

&15\% 
& 0.965  & 53.57 &0.9642  &  53.305&  0.9466&41.851 &0.9682 &55.171&0.9822  & 86.265\\

&20\% &  0.971 & 63.276 & 0.9601 &  51.724&0.9542  & 52.158& 0.9683 &54.125  & 0.9766 &84.267 \\

&25\% & 0.975  & 53.999 &0.9608& 53.02&0.9562  & 59.868&0.9742& 54.949 & 0.9805 &83.716 \\
\hline  
\multirow{5}{*}{10}&5\% 
&  0.974 & 86.641 &0.9728  &  84.273& 0.9702&69.721 & 0.9780&85.648 &0.9745  & 91.919\\

&10\% &  0.960 & 72.561
 & 0.9643 &  71.819&0.9512& 74.435& 0.9699&72.71 & 0.9773 &89.754 \\

&15\% & 0.964  & 56.111
 & 0.9615 &  52.925& 0.9479 &37.903 & 0.9663& 55.677& 0.9739 & 87.239\\

&20\% &  0.931 & 54.547
 & 0.9597 &  52.576&  0.9542& 56.327&0.9641 &55.584 &0.9829  &84.032 \\

&25\% &   0.956&56.626  &  0.9692&  52.155& 0.9534 &52.814 &0.9706 &54.114 & 0.9758 &89.977
 \\
\hline  
\multirow{5}{*}{15}&5\% 
& 0.966  & 86.321 & 0.9728 &  84.739&  0.9713& 71.413&0.9789 &86.332 &0.9800  &91.276 \\

&10\% 
& 0.960 & 71.473 &0.9634  &  70.516&  0.9515&75.897 & 0.9691&77.834 & 0.9739 &89.513 \\

&15\% 
&0.949 &62.773 &0.9597 &  57.298& 0.9476 &55.617 & 0.9654& 57.122& 0.9817 & 86.213\\

&20\% 
&0.940   & 56.86 &  0.9577&  53.013& 0.9535 &59.302&0.9617 &54.712 & 0.9773 & 85.412\\

&25\% &  0.948 &  57.709& 0.9674 &  52.405& 0.9529 &50.01 &0.9681 &58.111 & 0.9740 &91.213 \\
\hline

\end{tabular}
}
\end{table*}
\begin{table*}[h!]
\centering
\caption{Comparison of the proposed method \textit{GShield} under different safe rounds on the image dataset. Bold indicates best results.}
\label{tab:Imagedatset}
\renewcommand{\arraystretch}{1.2}
\resizebox{\textwidth}{!}{
\begin{tabular}{cccccccccccc}
\hline
\multirow{2}{*}{$SafeRound\downarrow$}&\multirow{2}{*}{$PMR\downarrow$}  
& \multicolumn{2}{c}{Krum} 
& \multicolumn{2}{c}{TMean} 
& \multicolumn{2}{c}{Median} 
& \multicolumn{2}{c}{FLAME} 
& \multicolumn{2}{c}{GShield} \\
\cline{3-12}
&& F1Score & SRecall & F1Score & SRecall & F1Score & SRecall & F1Score & SRecall & F1Score & SRecall \\
\hline  
\multicolumn{12}{c}{\textbf{\textit{FashionMNIST}}} \\
\hline
\multirow{5}{*}{5}&5\% 
& 0.854  & 94.456 &  0.831& 93.306 & 0.865 &94.958 &0.870 &95.554 &0.878 &95.922 \\

&10\% & 0.852  & 91.480 & 0.848&93.110 &0.864  &92.602  & 0.865& 92.438&0.876&93.614 \\

&15\% &  0.852 &91.030 &0.852  & 93.846 &  0.867&93.158 &0.871 &92.274 &0.872 &93.226 \\

&20\% &  0.856 &  86.678& 0.827 & 92.902 &0.866  & 90.432&0.865 & 88.274& 0.868&93.326 \\

&25\% &  0.837 & 80.516 &0.840  &88.750  & 0.847 & 83.500&0.851 &84.984 & 0.863&92.280 \\
\hline
\multirow{5}{*}{10}&5\%& 0.873  &  94.36&0.831 & 93.31 & 0.865 &95.00 &0.868 &95.55 &0.877 & 95.90\\

&10\% &  0.872 &  91.42& 0.848 &92.11  & 0.865 & 92.61& 0.864& 92.46&0.874 &93.71 \\

&15\% &  0.872 & 90.83&  0.852& 92.85 & 0.866 &93.22 &0.870 &92.47 &0.873 &93.56 \\

&20\% &  0.865& 85.93 &0.827  & 90.90 & 0.867 & 91.12& 0.866&89.89 & 0.871& 92.60\\

&25\% &0.856   &  80.16&0.840  &88.75  & 0.848 &85.28 &0.855 & 887.77&0.858 & 90.17\\
\hline
\multirow{5}{*}{15}&5\% & 0.875  & 95.64 & 0.833 &  97.24& 0.866 &95.35 &0.870 &96.06 & 0.878& 97.27\\

&10\% &  0.872 & 93.40 &  0.846&  93.29& 0.864 &93.16 & 0.864& 93.63&0.871 & 94.64\\

&15\% & 0.874  &  94.07&0.851  &  94.17& 0.868 & 94.36&0.869 & 94.28&0.872 &95.19 \\

&20\% &0.871  & 92.78&0.830 & 93.18 & 0.869 & 93.70 & 0.870& 93.45& 0.872& 94.53 \\

&25\% &   0.867&91.52  & 0.840 &90.02  & 0.856 &92.54 &0.859& 93.71& 0.866&93.98 \\
\hline
\multicolumn{12}{c}{\textbf{\textit{EMNIST}}} \\
\hline
\multirow{5}{*}{5}&5\% & 0.9808  &95.22  &0.9845  &  95.481&0.9601  & 95.711& 0.9881&96.975 &0.9851 &99.193 \\

&10\% &   0.9774& 94.382 &0.9812  & 94.142 & 0.9403 & 93.688&0.9883 & 95.471& 0.9842&98.891 \\

&15\% &  0.9745 & 88.793 & 0.9718 &88.918  & 0.9152 & 86.424& 0.9848&89.989 &0.9834& 99.064\\

&20\% &  0.9727 & 88.802 &  0.9671& 88.951 & 0.9016 & 88.218&0.9835 & 88.725& 0.9824&98.343 \\

&25\% & 0.9760  & 86.77 &0.9799  &85.046  & 0.8988 &85.731 &0.9832 &87.278  & 0.9839& 97.949\\
\hline
\multirow{5}{*}{10}&5\% &0.979  & 95.582&0.985   & 95.623 &  0.960  & 95.285& 0.988& 96.889& 0.984& 98.022\\

&10\% &  0.978  & 94.077 &  0.981 & 93.325 &  0.943  & 92.239& 0.988&95.391 &0.987 &95.93 \\

&15\% & 0.975  & 88.817 & 0.972  &89.379  &0.916 & 87.356&0.985 & 90.24&0.987 & 93.519\\

&20\% & 0.974   & 86.638 & 0.963& 88.334 & 0.905 &87.895 &0.983& 88.731& 0.990&91.57 \\

&25\% & 0.974&85.145 &0.963  &85.846  &0.909  &84.908 & 0.985&87.047 &0.983 & 91.903\\
\hline
\multirow{5}{*}{15}&5\% &0.979& 95.796  &0.976 & 95.231&0.974 & 95.374   & 0.988&96.894 & 0.984 &97.27 \\

&10\% &  0.973 &  88.597&  0.971& 94.352 &  0.960& 94.302& 0.987&95.373 &0.981 & 97.924\\

&15\% & 0.978  & 86.866 &0.948  & 89.585 & 0.882 &89.838 & 0.984&90.322 & 0.980& 93.743\\

&20\% & 0.978  & 86.866 & 0.925 & 87.088 &0.955  & 85.582&0.982 &89.024 &0.989 & 92.986\\

&25\% &  0.978 &  86.866&0.979  & 84.849 &0.954  &84.968 & 0.984& 86.741& 0.983&88.391 \\
\hline

\end{tabular}
}
\end{table*}

\subsection{Performance Analysis of GShield in Tabular Dataset}
In this section, we examine the robustness of the proposed GShield method under various safe zones configurations and compare its performance with that of the state-of-the-art method. To assess the sensitivity to the safe period, we configured $SafeRound$ to {5, 10, 15}. We used the same experimental setup for all state-of-the-art approaches. Specifically, no malicious participants were introduced until the prescribed $SafeRound$, after which malicious participants were added in subsequent rounds. Table \ref{tab:tabulardataset} presents the performance of our GShield framework compared to state-of-the-art methods on two tabular benchmark datasets under varying PMR attack ratios and different safe rounds for brevity. GShield consistently outperforms other methods across all datasets and attack scenarios. For CIC-Darkent2020, at lower safe rounds (e.g., 5 rounds), all approaches exhibit relatively low recall values due to the difficulty of detecting malicious updates in a highly non-IID environment. Nevertheless, the proposed GShield demonstrates effectiveness, achieving a 5.4\% higher recall than the best baseline (FLAME) at 10\% PMR, highlighting its enhanced capability to identify adversarial clients even during early aggregation phases.

As the number of safe rounds increases (e.g., 10--15), the overall recall of all methods improves since additional aggregation rounds provide better global context for anomaly detection. Importantly, GShield maintains its robustness and detection stability, achieving $90.55\%$ recall at 15 safe rounds and $10\%$ PMR. While the numerical gain (approximately $1.6\%$) over the next best approach may appear modest, it reflects a \textit{significant consistency advantage}, demonstrating that GShield sustains high detection accuracy even as the model converges and attack strength intensifies. This stable performance across varying conditions validates GShield's resilience and its adaptability to heterogeneous, non-IID federated learning environments.


For the CSE-CIC-IDS2018 dataset, we observed that GShield achieved approximately $64.884\% $ to $43.253\%$ higher performance for the $SRecall$ than other methods when $SafeRound$ was set to 5 with a PMR of $5\%$. In the case of a $safeRound$ of 5 with a PMR of $25\%$, the $SRecall$ for GShield showed an improvement of approximately $76\%$ compared with other methods, while all other approaches were closer to $0\%$. This result demonstrates GShield's ability to detect malicious participants effectively while maintaining the $SRecall$. Similarly, in this setting, GShield maintained an $F1Score$ of approximately $87\%$ across all cases, which was higher than the other approaches, whose scores ranged from $70\%$ to $72\%$. A similar trend was observed in other cases as well. The proposed GShield maintained the performance of the global model by effectively detecting both malicious and low-quality models during each round. Under the worst-case attack setting ($SafeRound = 15$ and $PMR = 15\%$), the $SRecall$ reached $91.213\%$, which is approximately $33\%$ to $41\%$ higher than that of other methods. Meanwhile, the global model achieved an $F1Score$ of $0.9740$.

\subsection{Performance Analysis of GShield in Image Datasets}

Table \ref{tab:Imagedatset} illustrates the performance of the proposed GShield method under different  $SafeRound$  and attack scenario settings. In the FashionMNIST dataset, for the $SRecall$ metric, GShield achieved approximately $2–3\%$ improvement over the competing methods for $ PMR=5\%$ and $SafeRound=5$.  Similarly, we observed that the $F1Score$ of the proposed GShield and FLAME remained almost identical. However, compared to other approaches, GShield improved approximately $2–4\%$. As the poisoning intensity increased, the $SRecall$ of all approaches declined significantly, although most methods attempted to maintain a stable $F1Score$. Notably, GShield still demonstrated an improvement of about $4–12\%$ when the $PMR$ increased to $25\%$, highlighting its robustness in handling higher levels of data poisoning.

Furthermore, as the $SafeRound$ increased, GShield consistently outperformed the other approaches. However, in most cases, FLAME's performance was comparable to GShield, primarily because FLAME is capable of distinguishing anomalous models even under a high proportion of malicious participants. Nevertheless, GShield outperformed all other methods in both metrics, achieving approximately $1–3\%$ higher performance in $SRecall$, while maintaining a comparable $F1Score$ under the worst-case attack scenario, where other methods exhibited significant degradation.

\subsection{Performance Analysis of GShield Effectiveness with Differential Privacy}

\begin{figure*}[h!]
\centering
\begin{subfigure}{0.3\textwidth}
    \includegraphics[width=0.9\linewidth]{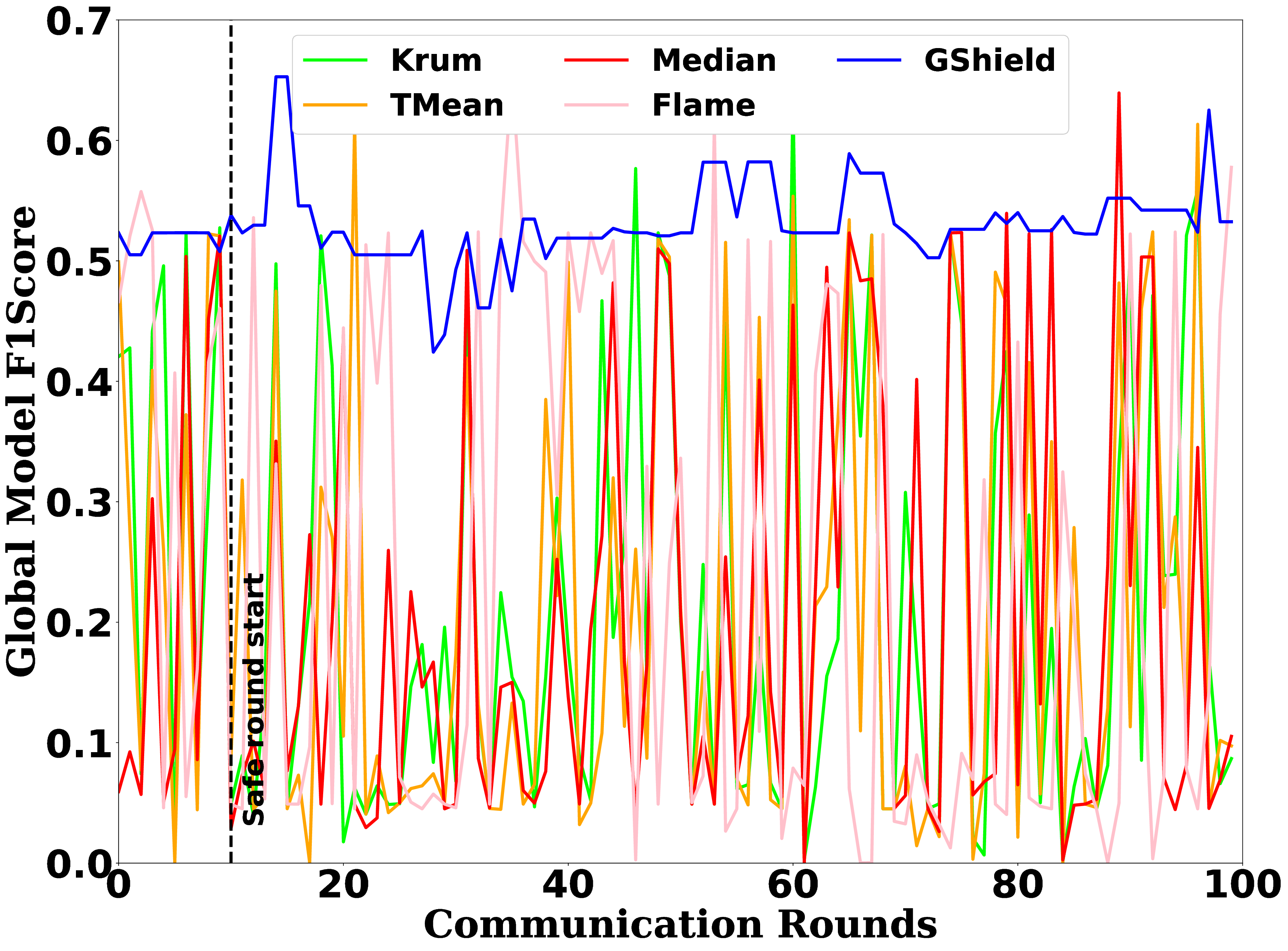} 
    \caption{CIC-Darknet2020}
    \label{fig:cicdarknet_noniid}
\end{subfigure}
\hfill
\begin{subfigure}{0.3\textwidth}
    \includegraphics[width=0.9\linewidth]{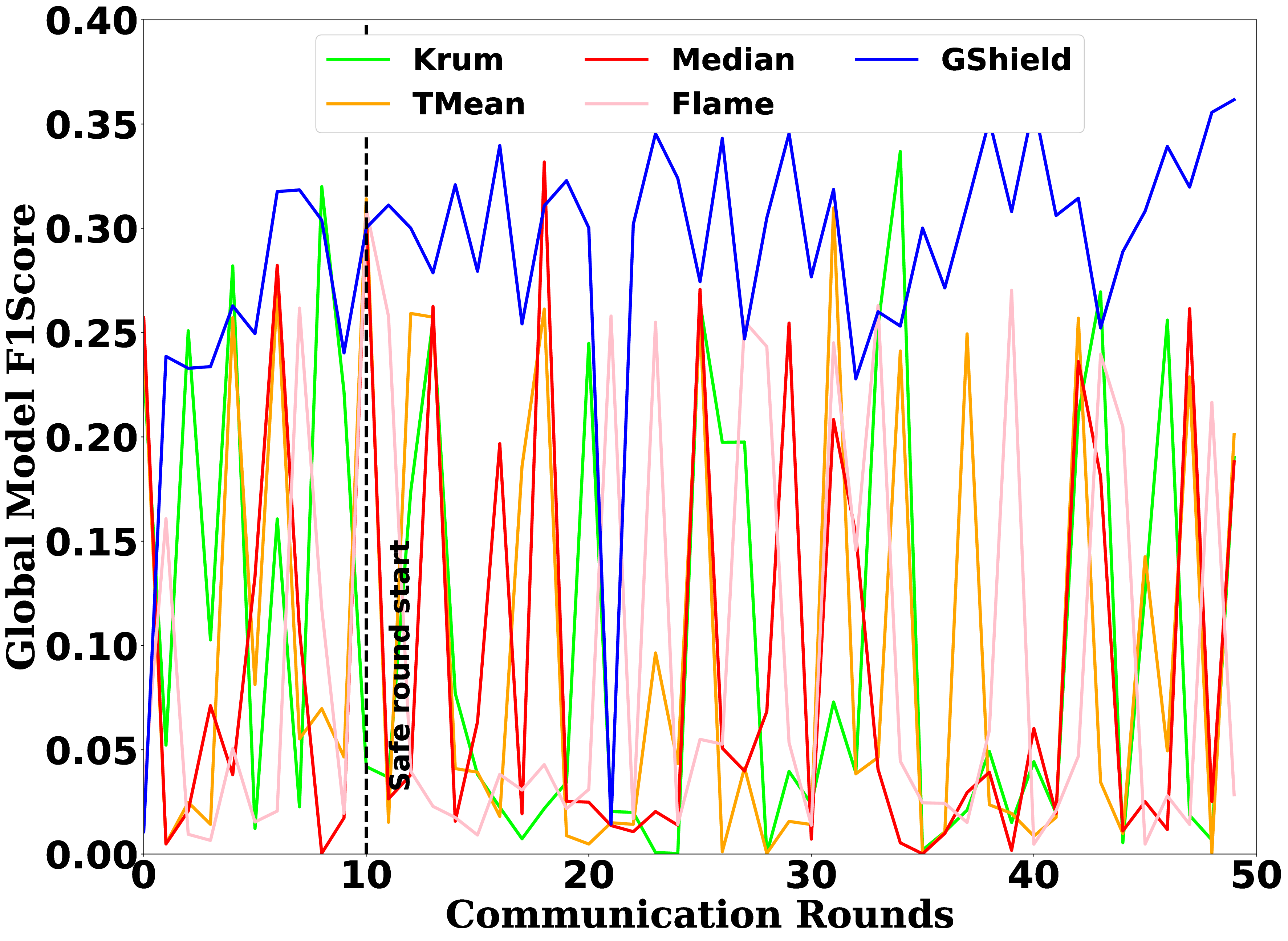} 
    \caption{CIC-IDS2018}
    \label{fig:cicids2018}
\end{subfigure}
\hfill
\begin{subfigure}{0.3\textwidth}
    \includegraphics[width=0.9\linewidth]{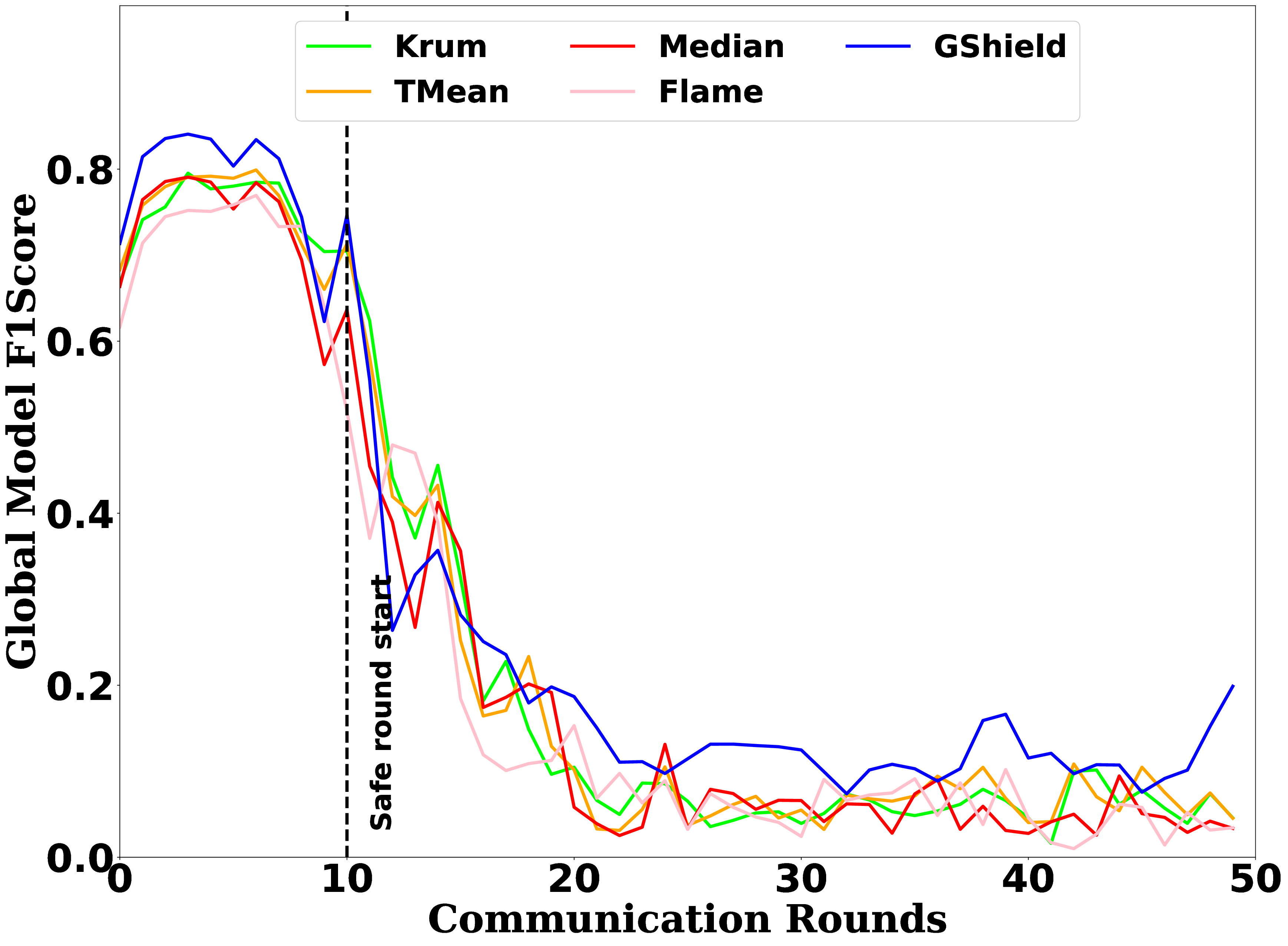} 
    \caption{FashionMNIST}
    \label{fig:cicids2018}
\end{subfigure}
\caption{Performance comparison demonstrating GShield’s robustness under Differential Privacy.}
\label{fig:DP_scenraio}
\end{figure*}

To further investigate the effectiveness of GShield when integrated with Differential Privacy (DP), we compared its performance against several state-of-the-art defense frameworks under the same attack scenarios. In this experiment, $25\%$ of the participating clients were designated as malicious, each poisoning $50\%$ of their local samples, and the SafeRound was set to $10$ to evaluate the system’s resilience under these conditions. The objective was to assess how well GShield preserves model accuracy and robustness while maintaining privacy protection. As shown in Figure \ref{fig:DP_scenraio}, GShield consistently outperformed existing methods in mitigating the impact of poisoning attacks, even under differential privacy constraints. This demonstrates that the incorporation of DP does not significantly compromise the model's learning capability. Moreover, GShield achieved faster convergence and maintained stable performance across different attack scenarios, highlighting its ability to balance both privacy preservation and defense robustness more effectively than existing approaches.

\subsection{Computational Overhead of GShield}

In this section, we discuss the computation overhead of the proposed approach against other state-of-the-art methods, whose results are visible in Table \ref{tab:computation_overhead}. The result illustrated that our proposed GShield has lower computation compared with Krum and FLAME. For the FashionMNIST dataset, GShield ($0.355$ seconds) is $80.9\%$ faster than FLAME ($1.859$ seconds) and $49.6\%$ faster than Krum ($0.705$ seconds). A similar trend is observed across other datasets. On CIC-Darknet2020, GShield reduces computation costs by $69.9\%$ compared to FLAME and by $67.7\% $compared to Krum.
Additionally, on the EMNIST dataset, GShield shows a $78.1\%$ improvement over FLAME. The TMean and Median have the lowest computational cost, ranging from $0.002$ to $0.038$ seconds. However, their simplicity comes from basic statistical aggregation methods that do not address adversarial manipulation. As a result, these approaches can be easily compromised by targeted poisoning strategies. In contrast, GShield provides robustness against malicious updates with only a slight increase in runtime compared with this approach. This ensures effective security in FL systems by filtering out malicious model updates without imposing significant latency on the aggregation process.

\begin{table*}[]
\centering

\caption{Comparison of GShield's computational time (in seconds) with the state-of-the-art approach.}
\label{tab:computation_overhead}
\begin{tabular}{lcccc}
\hline
\multirow{2}{*}{Method$\downarrow$}&\multicolumn{4}{c}{Dataset}\\
\cmidrule(lr){2-3}
\cmidrule(lr){4-5}
 &  CIC-Darknet2020 & CSE-CIC-IDS2018 & FashionMNIST&EMNIST\\ \hline
 Krum~\cite{blanchard2017machine} &0.372&0.216&0.705&0.659\\
 TMean~\cite{yin2018byzantine}&0.003&0.003  &0.020&0.038\\
Median~\cite{yin2018byzantine} &0.002&0.004  &0.019&0.031\\
FLAME~\cite{nguyen2022flame}&0.399&0.863&1.859  &1.143\\
GShield (proposed)&0.120& 0.138 &0.355&0.250\\

\hline
\end{tabular}

\end{table*}

\section{Conclusion}
\label{sec:conclusion}

In this paper, we introduced GShield, a robust server-side defense mechanism designed to mitigate targeted data poisoning attacks in Federated Learning, particularly under non-IID data distributions. By learning a statistical representation of benign client behavior through gradient clustering and Gaussian modeling, GShield effectively identifies and filters malicious and low-quality updates without requiring prior knowledge of adversaries or auxiliary datasets. Extensive experimental results demonstrate that GShield significantly improves model robustness and preserves high accuracy across both tabular and image datasets when compared to state-of-the-art defenses. 

We plan to extend our work in the future to handle more adaptive and colluding adversaries, explore its effectiveness against other attack types such as model replacement and backdoor attacks, and investigate dynamic or continual updating of the benign profile to further enhance robustness in long-running Federated Learning deployments.


\bibliographystyle{plain}
\bibliography{biblio}

@article{blanchard2017machine,
  title={Machine learning with adversaries: Byzantine tolerant gradient descent},
  author={Blanchard, Peva and El Mhamdi, El Mahdi and Guerraoui, Rachid and Stainer, Julien},
  journal={Advances in neural information processing systems},
  volume={30},
  year={2017}
}

@misc{cao2022fltrust,
      title={FLTrust: Byzantine-robust Federated Learning via Trust Bootstrapping}, 
      author={Xiaoyu Cao and Minghong Fang and Jia Liu and Neil Zhenqiang Gong},
      year={2022},
      eprint={2012.13995},
      archivePrefix={arXiv},
      primaryClass={cs.CR},
      url={https://arxiv.org/abs/2012.13995}, 
}

@article{chen2020zero,
  title={Zero knowledge clustering based adversarial mitigation in heterogeneous federated learning},
  author={Chen, Zheyi and Tian, Pu and Liao, Weixian and Yu, Wei},
  journal={IEEE Transactions on Network Science and Engineering},
  volume={8},
  number={2},
  pages={1070--1083},
  year={2020},
  publisher={IEEE}
}

@inproceedings{chelli2023fedguard,
  title={Fedguard: Selective parameter aggregation for poisoning attack mitigation in federated learning},
  author={Chelli, Melvin and Prigent, C{\'e}dric and Schubotz, Ren{\'e} and Costan, Alexandru and Antoniu, Gabriel and Cudennec, Lo{\"\i}c and Slusallek, Philipp},
  booktitle={2023 IEEE International Conference on Cluster Computing (CLUSTER)},
  pages={72--81},
  year={2023},
  organization={IEEE}
}

@misc{fu2021attack,
      title={Attack-Resistant Federated Learning with Residual-based Reweighting}, 
      author={Shuhao Fu and Chulin Xie and Bo Li and Qifeng Chen},
      year={2021},
      eprint={1912.11464},
      archivePrefix={arXiv},
      primaryClass={cs.LG},
      url={https://arxiv.org/abs/1912.11464}, 
}

@misc{fung2020mitigating,
      title={Mitigating Sybils in Federated Learning Poisoning}, 
      author={Clement Fung and Chris J. M. Yoon and Ivan Beschastnikh},
      year={2020},
      eprint={1808.04866},
      archivePrefix={arXiv},
      primaryClass={cs.LG},
      url={https://arxiv.org/abs/1808.04866}, 
}

@inproceedings{guerraoui2018hidden,
  title={The hidden vulnerability of distributed learning in byzantium},
  author={Guerraoui, Rachid and Rouault, S{\'e}bastien and others},
  booktitle={International conference on machine learning},
  pages={3521--3530},
  year={2018},
  organization={PMLR}
}

@article{hsu2019measuring,
  title={Measuring the effects of non-identical data distribution for federated visual classification},
  author={Hsu, Tzu-Ming Harry and Qi, Hang and Brown, Matthew},
  journal={arXiv preprint arXiv:1909.06335},
  year={2019}
}

@article{hernandez2025intrusion,
  title={Intrusion Detection based on Federated Learning: a systematic review},
  author={Hernandez-Ramos, Jose Luis and Karopoulos, Georgios and Chatzoglou, Efstratios and Kouliaridis, Vasileios and Marmol, Enrique and Gonzalez-Vidal, Aurora and Kambourakis, Georgios},
  journal={ACM Computing Surveys},
  volume={57},
  number={12},
  pages={1--65},
  year={2025},
  publisher={ACM New York, NY}
}

@inproceedings{liu2022backdoor,
  title={Backdoor defense with machine unlearning},
  author={Liu, Yang and Fan, Mingyuan and Chen, Cen and Liu, Ximeng and Ma, Zhuo and Wang, Li and Ma, Jianfeng},
  booktitle={IEEE INFOCOM 2022-IEEE conference on computer communications},
  pages={280--289},
  year={2022},
  organization={IEEE}
}

@article{ma2022shieldfl,
  title={ShieldFL: Mitigating model poisoning attacks in privacy-preserving federated learning},
  author={Ma, Zhuoran and Ma, Jianfeng and Miao, Yinbin and Li, Yingjiu and Deng, Robert H},
  journal={IEEE Transactions on Information Forensics and Security},
  volume={17},
  pages={1639--1654},
  year={2022},
  publisher={IEEE}
}

@inproceedings{mcmahan2017communication,
  title={Communication-Efficient Learning of Deep Networks from Decentralized Data},
  author={McMahan, H Brendan and Moore, Eider and Ramage, Daniel and Hampson, Seth and Aguera y Arcas, Blaise},
  booktitle={Proceedings of the 20th International Conference on Artificial Intelligence and Statistics (AISTATS)},
  volume={54},
  pages={1273--1282},
  year={2017},
  organization={PMLR}
}

@article{munoz2019byzantine,
  title={Byzantine-robust federated machine learning through adaptive model averaging},
  author={Mu{\~n}oz-Gonz{\'a}lez, Luis and Co, Kenneth T and Lupu, Emil C},
  journal={arXiv preprint arXiv:1909.05125},
  year={2019}
}

@article{nicolazzo2025secure,
  title={How Secure is Forgetting? Linking Machine Unlearning to Machine Learning Attacks},
  author={Nicolazzo, Serena and Nocera, Antonino and others},
  journal={arXiv preprint arXiv:2503.20257},
  year={2025}
}

@inproceedings{nguyen2022flame,
  title={$\{$FLAME$\}$: Taming backdoors in federated learning},
  author={Nguyen, Thien Duc and Rieger, Phillip and Chen, Huili and Yalame, Hossein and M{\"o}llering, Helen and Fereidooni, Hossein and Marchal, Samuel and Miettinen, Markus and Mirhoseini, Azalia and Zeitouni, Shaza and others},
  booktitle={31st USENIX security symposium (USENIX Security 22)},
  pages={1415--1432},
  year={2022}
}

@article{pillutla2022robust,
  title={Robust aggregation for federated learning},
  author={Pillutla, Krishna and Kakade, Sham M and Harchaoui, Zaid},
  journal={IEEE Transactions on Signal Processing},
  volume={70},
  pages={1142--1154},
  year={2022},
  publisher={IEEE}
}

@article{sameera2024privacy,
  title={Privacy-preserving in Blockchain-based Federated Learning systems},
  author={Sameera, KM and Nicolazzo, Serena and Arazzi, Marco and Nocera, Antonino and KA, Rafidha Rehiman and Vinod, P and Conti, Mauro},
  journal={Computer Communications},
  year={2024},
  publisher={Elsevier}
}

@article{sameera2025federated,
  title={Federated Learning: An Overview of Attacks and Defense Methods},
  author={Sameera, KM and Arikkat, Dincy R and Vinod, P and Rafidha, Rehiman KA and Aneez, Azin and Conti, Mauro},
  journal={Machine Learning, Deep Learning and AI for Cybersecurity},
  pages={393--431},
  year={2025},
  publisher={Springer}
}

@misc{sun2019really,
      title={Can You Really Backdoor Federated Learning?}, 
      author={Ziteng Sun and Peter Kairouz and Ananda Theertha Suresh and H. Brendan McMahan},
      year={2019},
      eprint={1911.07963},
      archivePrefix={arXiv},
      primaryClass={cs.LG},
      url={https://arxiv.org/abs/1911.07963}, 
}

@article{thein2024personalized,
  title={Personalized federated learning-based intrusion detection system: Poisoning attack and defense},
  author={Thein, Thin Tharaphe and Shiraishi, Yoshiaki and Morii, Masakatu},
  journal={Future Generation Computer Systems},
  volume={153},
  pages={182--192},
  year={2024},
  publisher={Elsevier}
}

@inproceedings{tolpegin2020data,
  title={Data poisoning attacks against federated learning systems},
  author={Tolpegin, Vale and Truex, Stacey and Gursoy, Mehmet Emre and Liu, Ling},
  booktitle={European symposium on research in computer security},
  pages={480--501},
  year={2020},
  organization={Springer}
}

@article{yang2025enhanced,
  title={Enhanced model poisoning attack and multi-strategy defense in federated learning},
  author={Yang, Li and Miao, Yinbin and Liu, Ziteng and Liu, Zhiquan and Li, Xinghua and Kuang, Da and Li, Hongwei and Deng, Robert H},
  journal={IEEE Transactions on Information Forensics and Security},
  year={2025},
  publisher={IEEE}
}

@inproceedings{yin2018byzantine,
  title={Byzantine-robust distributed learning: Towards optimal statistical rates},
  author={Yin, Dong and Chen, Yudong and Kannan, Ramchandran and Bartlett, Peter},
  booktitle={International conference on machine learning},
  pages={5650--5659},
  year={2018},
  organization={Pmlr}
}

@article{wang2025federated,
  title={Federated learning framework based on trimmed mean aggregation rules},
  author={Wang, Tianxiang and Zheng, Zhonglong and Lin, Feilong},
  journal={Expert Systems with Applications},
  pages={126354},
  year={2025},
  publisher={Elsevier}
}

@article{zhao2021sear,
  title={Sear: Secure and efficient aggregation for byzantine-robust federated learning},
  author={Zhao, Lingchen and Jiang, Jianlin and Feng, Bo and Wang, Qian and Shen, Chao and Li, Qi},
  journal={IEEE Transactions on Dependable and Secure Computing},
  volume={19},
  number={5},
  pages={3329--3342},
  year={2021},
  publisher={IEEE}
}

\end{document}